\documentclass[preprint,12pt]{elsarticle}

\usepackage{amssymb}
\usepackage{array}
\usepackage{graphicx}
\usepackage{booktabs}
\usepackage[table]{xcolor}
\usepackage{lscape}
\usepackage{makecell}
\usepackage{tabularx}
\usepackage{float}
\usepackage{adjustbox}
\usepackage{url}

\newcommand{\wrappedcell}[1]{%
    \makecell{\begin{minipage}[t]{\linewidth}#1\end{minipage}}%
}
\newtheorem{definition}{Definition}
\newcolumntype{T}[1]{>{\raggedright\arraybackslash\setlength{\topsep}{0pt}\setlength{\partopsep}{0pt}}p{#1}}
\renewcommand{\arraystretch}{1.1}

\journal{Journal of Information Security and Applications}

\begin{document}

\begin{frontmatter}



\title{A Descriptive Model for Modelling Attacker Decision-Making in Cyber-Deception.}


\affiliation[inst1]{organization={Curtin University},
            addressline={Kent Street}, 
            city={Bentley},
            postcode={6102}, 
            state={WA},
            country={Australia}}

\affiliation[inst2]{organization={Edith Cowan University},
            addressline={270 Joondalup Drive}, 
            city={Joondalup},
            postcode={6027}, 
            state={WA},
            country={Australia}}

\affiliation[inst3]{organization={University of Wollongong},
            addressline={Northfields Ave}, 
            city={Wollongong},
            postcode={2500}, 
            state={NSW},
            country={Australia}}

\author[inst1,inst2]{B.R. Turner}

\author[inst3,inst2]{O. Guidetti}

\author[inst1]{N.M. Karie}

\author[inst1]{R. Ryan}

\author[inst1]{Y. Yan}

\begin{abstract}
Cyber-deception has become a critical strategy in cybersecurity, influencing adversarial decision-making through controlled misinformation, uncertainty, and misdirection. While existing models, including game-theoretic, Bayesian, Markov decision processes, and reinforcement learning approaches, have provided insight into how deception influences adversarial behaviour, these approaches primarily assume that an attacker has already committed to engagement. Although these models provide valuable insights into deception strategies, they often overlook the cognitive and perceptual factors influencing an attacker's initial decision to engage or disengage.

This paper introduces a descriptive model that accounts for the psychological and strategic elements that affect an attacker’s decision-making process. The model defines five core components: belief $(B)$, scepticism $(S)$, deception fidelity $(D)$, reconnaissance $(R)$, and experience $(E)$ which interact to describe how adversaries evaluate deception and determine whether to proceed with engagement. The model offers a systematic approach to analysing engagement decisions in cyber-deception scenarios by structuring these elements within a flexible framework.

A series of Capture the Flag experiments will be conducted to evaluate this model, which incorporates varying levels of deception. These experiments will integrate behavioural observations and biometric indicators to provide a multidimensional analysis of adversarial responses to deception. The findings will contribute to a deeper understanding of the factors influencing engagement decisions and refine the model's applicability to real-world cyber-deception environments.

The proposed model enhances theoretical and applied understandings of deception in cybersecurity. The findings of this study will provide a foundation for refining cyber-deception strategies by integrating cognitive realism into adversarial decision-making processes. By addressing the gap in existing models that assume engagement, the proposed model enhances theoretical and applied understandings of deception in cybersecurity.
\end{abstract}




\end{frontmatter}


\section{Introduction}
\label{Introduction}
Cyber-deception has become a fundamental aspect of cybersecurity strategy, enabling defenders to manipulate adversarial decision-making by introducing uncertainty, misdirection, and false information into digital environments. As cyber threats evolve, deception techniques have been increasingly adopted to disrupt attackers’ situational awareness, delay malicious activities, and force adversaries to expend additional resources. However, understanding how attackers decide to engage with deceptive elements remains an open research challenge, requiring a structured approach to adversarial decision-making.

Existing cyber-deception models, including game-theoretic, Bayesian, Markov decision processes (MDPs), and reinforcement learning approaches, provide valuable insights into how deception influences adversary behaviour. However, many of these models assume that an attacker has already committed to engagement, focusing on optimising deception strategies for after the decision to engage has been made, rather than examining the factors that lead to the decision to attack. In reality, adversaries rely on a combination of rational evaluation, cognitive biases, and heuristic reasoning when assessing potential targets, meaning that factors beyond cost-benefit calculations influence engagement decisions.

The proposed model in this paper addresses this gap by introducing a descriptive framework that accounts for the psychological and strategic elements influencing an attacker’s engagement decision. The model defines five core components—belief (B), scepticism (S), deception fidelity (D), reconnaissance (R), and experience (E)—which interact to describe the decision-making process leading to engagement. Unlike existing models that treat deception as a static interaction, this approach recognises the fluid nature of adversarial decision-making and the evolving perception of trust, uncertainty, and deception fidelity over time.

To evaluate this model, a structured series of Capture the Flag (CTF) experiments will be conducted, measuring how attackers respond to varying levels of deception. These experiments will incorporate both behavioural observations and biometric indicators to provide a multi-faceted analysis of adversarial decision-making. The results will provide empirical insights into how deception influences engagement and inform further refinements of the model.

\section{Literature Review}
Research into cyber deception has increasingly adopted formal models to represent attacker decision-making processes. Mathematical and computational frameworks provide a structured means of examining how attackers perceive, reason, and act within deceptive environments. Among these, Bayesian inference, Markov decision processes, and reinforcement learning approaches have received particular attention, each offering unique insights into the mechanisms of attacker cognition and adaptation. While their mathematical foundations differ, these models share a common goal: to approximate rational behaviour under uncertainty and to identify the conditions under which deception may succeed or fail.

Bayesian models represent attacker decisions as probabilistic reasoning under incomplete information. As new evidence becomes available, prior beliefs are updated to produce posterior probabilities that guide subsequent actions. This formulation captures the logic of belief revision and is particularly useful in settings where the attacker must infer the authenticity of targets from partial or ambiguous observations. However, while Bayesian reasoning embodies the principle of belief updating, it generally abstracts from the qualitative aspects of belief confidence and scepticism that influence engagement decisions. The present work therefore extends, rather than replaces, Bayesian perspectives by modelling these cognitive dimensions explicitly.

Markov decision processes (MDPs) and their derivatives express attacker behaviour as a sequence of probabilistic state transitions governed by defined rules of motion. Such models are effective for representing the temporal and adaptive nature of attack progression, particularly where system responses can be discretised into states. Yet, MDPs typically describe transitions without representing the cognitive mechanisms that motivate them. The model proposed here complements this tradition by incorporating cognitive variables—belief, scepticism, deception fidelity, experience, and reconnaissance—that explain why an attacker may choose one transition path over another.

Reinforcement learning (RL) approaches similarly represent adaptation through iterative feedback, allowing an agent to learn from environmental rewards and penalties. These models have proven effective in simulating red-team and blue-team dynamics, and in identifying strategies that maximise utility in competitive scenarios. However, they focus primarily on outcome optimisation rather than the cognitive reasoning that precedes behavioural change. The present model approaches the same problem from a complementary direction, defining engagement as a function of perceived authenticity, familiarity, and effort. In doing so, it expresses cognitive antecedents in a form that can later interface with data-driven RL or Bayesian frameworks for empirical calibration.

Collectively, these prior approaches provide the empirical and theoretical basis for understanding deception within cyber operations. The model developed in this work builds upon that foundation by prioritising interpretability and conceptual clarity. It aims to describe how cognitive factors, rather than solely probabilistic or state-based dynamics, shape attacker engagement decisions. This perspective aligns behavioural modelling with the psychological context in which deception operates, offering an abstract yet adaptable framework for subsequent empirical investigation.

A comparative summary of representative modelling approaches is presented in Table~\ref{tab:model_comparison}. 
The table highlights the primary characteristics, modelling focus, and contextual relevance of existing frameworks that inform the development of the present model. 
This summary provides a concise reference for understanding how prior approaches contribute to the broader study of attacker decision-making in deceptive environments.

\begin{table}[H]
    \centering
    { 
      \scriptsize
      \renewcommand{\arraystretch}{1.2}
      \begin{tabularx}{\linewidth}{|
          >{\raggedright\arraybackslash}X|
          >{\raggedright\arraybackslash}X|
          >{\raggedright\arraybackslash}X|
          >{\raggedright\arraybackslash}X|}
        \hline
        \textbf{Model \& Approach} & \textbf{Mathematical Foundation} & \textbf{Handling of Deception} & \textbf{Incorporation of Psychological Factors} \\ \hline
        \wrappedcell{Proposed model} &
        \wrappedcell{- Linear deception function with heuristic weighting \\ - $(B-S)-R(E-D)$: integrates belief ($B$), scepticism ($S$), experience ($E$), deception fidelity ($D$), and reconnaissance ($R$)} &
        \wrappedcell{- Dynamic: Outcome iterates based on real-time feedback (e.g. new reconnaissance data) \\ - Bounded reconnaissance ($R$) scales the impact of deception vs. experience} &
        \wrappedcell{- Explicitly separates trust (belief vs. scepticism) and experience \\ - Emphasises cognitive biases and heuristics \\ - Models the interplay of psychological and rational factors in one formula} \\ \hline
        \wrappedcell{Game Theory \\ (Thakoor, 2019)} &
        \wrappedcell{- Equilibrium concepts (e.g. Nash, Bayesian) \\ - Assumes rational adversarial play} &
        \wrappedcell{- Typically static: Solutions revolve around precomputed equilibrium strategies \\ - Adversaries do not continuously update or adapt beyond the equilibrium logic} &
        \wrappedcell{- Rational-choice assumption overlooks emotional or heuristic influences \\ - Does not explicitly model trust erosion, moral considerations, or evolving scepticism} \\ \hline
        \wrappedcell{Bayesian Models \\ (Carroll, 2010; \\ Huang, 2018)} &
        \wrappedcell{- Bayesian inference and belief updating \\ - Perfect Bayesian Equilibrium in repeated or signalling scenarios} &
        \wrappedcell{- Belief manipulation is central, but the fundamental deception strategy can remain static \\ - Emphasises how signals change posterior beliefs, not how deception incentives shift over time} &
        \wrappedcell{- Presumes near-perfect Bayesian reasoning \\ - Often excludes real-world cognitive biases unless artificially introduced as “noise” \\ - Lacks explicit variables for trust or scepticism} \\ \hline
        \wrappedcell{MDPs \\ (Abri, 2022; \\ Ferrer-Mestres, 2020; \\ Grand-Clément, 2021)} &
        \wrappedcell{- State-action models governed by Bellman equations \\ - Probabilistic transitions that maximise expected long-term reward} &
        \wrappedcell{- Dynamic: Attackers can shift between cooperation and deception over sequential states \\ - Deception is orchestrated by reward structures and defined transitions} &
        \wrappedcell{- Trust, suspicion, or heuristic biases must be hard-coded into the state or transition structure \\ - Not inherently aligned with psychological realism (no direct representation of scepticism)} \\ \hline
        \wrappedcell{Reinforcement Learning \\ (He, 2023; \\ Stember, 2021)} &
        \wrappedcell{- Extends MDPs with policy iteration (Q-learning, PPO, etc.) \\ - Iterative improvement through trial-and-error experience} &
        \wrappedcell{- Fully dynamic: Agents learn deception strategies over multiple episodes \\ - Emergent deception can be highly adaptive but also potentially opaque} &
        \wrappedcell{- No explicit trust mechanism unless manually incorporated \\ - Performance depends on data fidelity; if training does not capture human biases, results may not translate to actual adversaries} \\ \hline
      \end{tabularx}
    } 
    \caption{Summary of representative modelling approaches relevant to cyber deception research.}
    \label{tab:model_comparison}
\end{table}

\section {Proposed Model}
The terminology and conceptual frameworks of cyber-deception owe much of their lineage to foundational principles established in military contexts. Deception in warfare has long been a cornerstone of strategic and operational planning, aimed at manipulating adversaries’ perceptions and actions. By borrowing and adapting these principles, the domain of cyber-deception has developed a lexicon and set of methodologies that reflect its military origins while addressing the unique challenges of the cyber-domain. 
A central theme in military deception is the exploitation of adversarial belief systems to manipulate perceptions and decision-making. Hood\cite{Hood2023} describes the military use of narratives that resonate with an opponent’s existing beliefs, fostering scepticism toward their intelligence. This principle is mirrored in cyber-deception strategies, which leverage realistic simulations of systems or vulnerabilities to mislead attackers into engaging decoy targets. Edwards\cite{Edwards2004} expands on this concept by illustrating how deception creates illusions of strength in military operations—a tactic that parallels the use of high-fidelity decoys in cybersecurity to project the appearance of robust defences.
The experiential dimension of deception is another area where military origins influence cyber-deception. Black and Reid\cite{Black2020} argue that the ability of military personnel to recognise and counter deception is shaped by accumulated experience, a concept adapted into cyber-defence as organisations refine their strategies through interactions with sophisticated adversaries. Similarly, Knobloch et al.\cite{Knobloch2020} highlight the role of relational uncertainty in military contexts, a dynamic that cyber-deception exploits to sow doubt in attackers about the reliability of their reconnaissance and decision-making.
The design and implementation of cyber-deception strategies also draw heavily from military methodologies. The emphasis on detailed and coherent planning, as outlined by Benke et al.\cite{Benke2021}, reflects military doctrines that integrate deception into broader strategic frameworks. Grohe\cite{Grohe2007} reinforces this alignment, describing the military practice of creating confusion through conflicting information, which directly informs cyber-deception techniques like redirection shields and honeypots.
The qualities of belief, scepticism, experience, and deception fidelity are central to deception strategies in both military and cyber-domains. Belief and scepticism operate as psychological levers, manipulated to create confidence in false information or doubt in legitimate intelligence. In military contexts, this might involve crafting narratives to mislead opponents, while in cyber-security, it manifests as designing deceptive systems that exploit attackers’ cognitive biases. Experience shapes how adversaries and defenders alike interpret and respond to deception, whether through battlefield engagements or cyber-intrusions. Finally, deception fidelity, the degree to which a deceptive tactic mimics reality—determines the effectiveness of these strategies. High-fidelity deception, exemplified by lifelike decoys in military ambushes or realistic honeypots in cyber-defence, increases the likelihood of adversarial engagement and the extraction of valuable intelligence. These fundamental components of the decision-making process involved in assessing whether to engage a target will serve as the foundation for the proposed mathematical model.

\subsection{Methodology (Model Development Process)}
\subsubsection{Mathematical Representation}
\label{Definitions}
We introduce a mathematical model for understanding an attacker's decision-making process in choosing whether to attack a target in a cyber-attack, or not. This model provides a descriptive representation of attacker decision-making during standard deception decision-making processes, including TTD (Transitional Target Defence) and TTP (tactics, techniques, and procedures) scenarios. This model highlights the role of the main components involved in deciding to attack a target,
belief, deception fidelity, scepticism, experience, and reconnaissance in shaping engagement decisions. These models contribute to a deeper understanding of deception strategies in cybersecurity, with potential applications in threat detection and mitigation.

\subsubsection{Linear Model Justification}
The linear structure adopted for this model was selected for clarity, interpretability, and modularity. A linear formulation provides an interpretable structure in which the relative influence of each contributing variable, including belief ($B$), scepticism ($S$), deception fidelity ($D$), experience ($E$), and reconnaissance effort ($R$), can be examined directly while avoiding additional non-linear dependencies that would complicate interpretation at this stage of model development. The primary objective of this stage of development is to establish an interpretable representation of the relationship between cognitive and contextual factors that influence attacker decision-making.

Although non-linear models such as Bayesian, Markov, or reinforcement learning approaches may capture complex adaptive behaviour, they typically require large empirical datasets for calibration and often produce outcomes that are less interpretable in the absence of such data. In contrast, a linear model allows each variable’s contribution to be examined independently, facilitates parameter sensitivity analysis, and supports extension to more complex formulations when suitable empirical data become available. This approach therefore represents a deliberate methodological decision to construct a model that is analytically tractable, empirically scalable, and conceptually clear in representing the mechanisms of engagement within deceptive environments.

\subsubsection{Components and concepts}
An attacker generally surveys a network for a target of interest before selecting and engaging. The likelihood of engagement reflects a combination of cognitive and contextual indicators, such as value or ease of exploitability (i.e., the value the attacker believes the target may have). Surveying may also reveal potential dangers, such as possible intrusion detection systems (IDS), firewalls, potential deceptive devices, or even a suspicious network topology. Indicators of danger may raise the degree of scepticism or disbelief that the attacker has that the target is what it appears to be. We argue that an attacker will engage with a target when their belief level ($B$) exceeds their scepticism level ($S$).

\begin{definition}[Belief]
    Belief refers to an attacker accepting a given assertion as true or valid. In the context of this paper, this translates to the belief that the target holds enough value to be exploited. 
\end{definition}

\begin{definition}[Scepticism]
    Scepticism is a critical and cautious approach to evaluating and responding to digital security. It involves questioning the security measures, claims, technologies, and practices in the cybersecurity field to protect computer systems, networks, and data. It also encompasses an attacker's doubt that the target host is not what it presents itself as.
\end{definition}
Scepticism in cyber-security involves a critical and questioning approach to all aspects of digital security, aiming to uncover vulnerabilities, expose weaknesses, and ensure that effective security measures are in place to protect against cyber-threats. Maintaining the integrity and trustworthiness of digital systems and data is vital. In the context of this paper, scepticism refers to the level at which the attacker believes that the target is not what it presents itself to be. 

\begin{definition}[Reconnaissance]
Reconnaissance is the information-gathering process for making informed inferences. It involves gathering information and intelligence about a target system, network, or organisation, either directly or indirectly. The primary purpose of reconnaissance is to collect data to evaluate a target and, if the target is deemed of value, proceed to the next phase of planning and executing a successful cyber-attack. It is essentially the cyber equivalent of scouting or gathering information before launching an attack.
\end{definition}
Reconnaissance is a critical phase of a cyber-attack because it significantly increases the chances of a successful breach. Organisations and cybersecurity professionals invest in counter-reconnaissance measures to detect and thwart reconnaissance attempts before they can progress to more harmful stages of an attack.

The information gathered during reconnaissance is valuable for several reasons:
\begin{itemize}
    \item helps attackers identify potential vulnerabilities and weaknesses in the target's defences.
    \item allows attackers to choose the most effective attack vectors and tools.
    \item provides insight into the target's network architecture, which aids in planning the attack.
    \item helps attackers create convincing phishing or social engineering attacks by tailoring them to the target's characteristics.
\end{itemize}

\begin{definition}[Deception]
    Deception refers to intentionally misleading someone or causing them to believe something untrue, which involves creating a false impression or conveying false information to manipulate perceptions, beliefs, or actions. 
    In the context of this paper, deception is an umbrella term for the techniques that deceptive devices can use to create the perception in an observer that the device is what it purports to be. This includes, but is not limited to, techniques such as dazzling, masking, inventing, and repackaging. 
\end{definition}
Deception takes various forms, including verbal, non-verbal, or digital, and it often seeks to gain an advantage, conceal information, or achieve a specific outcome. Deception can occur in interpersonal communication, espionage, cyber-security, warfare, and many other contexts where information and perception play a crucial role. Deception represents how well the deceptive device uses and implements deceptive techniques such as mimicry, invention, imitation, dazzling, etc. 

\begin{definition}[Experience]
    Experience is the accumulation of knowledge and skills related to deceptive devices, developed through exposure to three sources: education, specialised skill acquisition, and practical exposure and interaction.  
\end{definition}
Experience with deceptive devices allows an attacker to recognise previously encountered deceptive techniques. Lower levels of experience will allow an attacker to recognise default configurations of deceptive tools or unusual network topologies. As experience with deceptive devices increases, encountered deceptive techniques become abstracted so that new and novel deceptive approaches are able to be recognised.

\subsection{Proposed Model}
\subsection{Proposed Model Derivation}
The foundational premise is that once belief outweighs scepticism, an attacker will engage with the targeted host, and the process towards exploitation will begin. 
\begin{equation} (B-S) \end{equation} \\
When examining the interaction of experience and deception, we postulate that when experience is greater than the novelty and depth of the deceptive techniques employed by the deceptive device, then experience will inform the attacker, to some degree, that they should modify their initial assessment of the target host away from engagement. Conversely, if the deceptive techniques are previously unseen or are of significant enough detail, then the deception is likely to misdirect the attacker and modify the initial assessment of the target host in favour of engagement. \begin{equation}
    (E-D)
\end{equation}
\\
Finally, the application of experience and the effect of deceptive techniques are impossible to apply if the target host is not examined in some way, either directly or indirectly. It is impossible to be deceived by something, before engagement, that is not engaged with, and it is unlikely that previous patterns or techniques will be recognised if those patterns are not observed. As such, reconnaissance is a limiting factor determining how well such features are observed and how significant the impact experience and deception has on the engagement outcome. Reconnaissance is not an amplifier of the experience-deception dynamic and, as such, is limited to the range of 0 to 1.  \\ 
\begin{equation}
    R(E-D)
\end{equation}
\centerline{Such that: $R \subseteq \mathbb R$ such that $\forall r \in:r$ $\forall [0,1]$}

\subsubsection{The Proposed Model of Deception}
Once the original model was deemed a failure we stepped back and examined the components. Equation 1 and 3 were working as expected but the impulse to multiply them was in error. Examining the components again, keeping in mind the previous models errors, the relationship would appear more akin to:
\begin{equation}
    (B-S) - R(E-D)
\end{equation}
By subtracting the reconnaissance, when experience is greater than deception, the result becomes more negative. This result demonstrates that when the attacker's experience outweighs the efforts of the deceptive techniques used within the deceptive device, the attacker will become more weary of engaging with a deceptive device regardless of the initial belief value. 
When deception is greater than experience, this produces a negative value to be subtracted from the belief value, becoming a positive addition to the belief value, indicating an increased likelihood of engaging with the deceptive device. Thus, this result demonstrates that when the deceptive techniques are more persuasive than the attacker's experience would allow them to readily identify, the attacker becomes more likely to engage with the deceptive device independent of the original value of belief.
Given this relationship, the model could also be rewritten as:
\begin{equation}
    (B-S) + R(D-E)
\end{equation}
In both of these versions of the model, the outcome is the same. The more positive the result, the more likely engagement is to occur and the more negative, the less likely the attacker is to engage with the deceptive device. When these models are run through the test data, the output from the model matches the expected results. 

\subsection{Interpreting the Model Output}
The decision threshold in the model is given by the expression:
\begin{equation} (B-S)-R(E-D)  
\end{equation}
In this formulation, a high, positive value indicates a strong likelihood that an attacker will engage with a deceptive device. Conversely, a negative value suggests that the attacker is unlikely to engage, as their scepticism or accumulated experience outweighs the novelty of the deception. Values that are closer to zero indicate increasing uncertainty, meaning the decision to engage or disengage is finely balanced.

In summary, the proposed deception model offers an abstract yet robust approach that elucidates the intricate process of making an attack decision. By integrating key variables, belief ($B$), scepticism ($S$), experience ($E$), deception fidelity ($D$), and reconnaissance ($R$), the model not only fills important gaps in existing approaches but also reveals how attackers weigh various cognitive and rational factors when deciding whether to engage with a target. The computed threshold from the model, expressed as
\begin{equation}
    (B-S)-R(E-D)
\end{equation},
provides a quantifiable metric where high positive values indicate a strong likelihood of engagement, negative values suggest disengagement, and intermediate values reflect uncertainty. This delineation helps to unpack the decision-making process that leads to the intention to attack. this model demonstrates how an attacker’s perception of a target’s value, adjusted by their inherent scepticism and influenced by their past experiences relative to the sophistication of the deception encountered, ultimately drives this decision. Such an understanding helps understand adaptive cyber-defence strategies that can manipulate deceptive cues and dynamically influence attacker behaviour, thereby contributing to more resilient and responsive operational defence frameworks.

\section{Model Integration with Cyber-Attack Decision-Making Frameworks}
The descriptive model presented in this work is intended to complement existing cyber-deception and cyber-attack decision-making frameworks by providing a cognitive layer that represents attacker perception prior to behavioural adaptation. Computational approaches such as Bayesian inference, Markov decision processes, and reinforcement learning offer mechanisms for modelling uncertainty, sequential decision-making, and strategic adaptation, yet they typically do not represent the cognitive factors that shape an attacker's initial assessment of a potential target. The present model addresses this gap by expressing belief, scepticism, deception fidelity, experience, and reconnaissance in an interpretable form that can be integrated into broader defensive workflows. The engagement output produced by the model, expressed as a bounded probabilistic measure, can inform higher-level decision-making processes by estimating the likelihood that an attacker will interact with a deceptive artifact under specific contextual conditions. This representation provides a basis for integrating cognitive considerations with algorithmic approaches, supporting both conceptual analysis and the potential development of hybrid systems that incorporate empirical or data-driven methods as they become available.

\subsection{Model Evaluation}
The proposed model will be evaluated through a series of structured Capture the Flag (CTF) events designed to assess its applicability in real-world cyber-deception scenarios. These experiments will incorporate varying levels of deception, allowing for an analysis of how attackers perceive, react to, and engage with deceptive elements. By systematically altering deception techniques across different experimental conditions, the evaluation process will provide empirical data to examine how belief, scepticism, deception fidelity, reconnaissance, and experience influence engagement decisions.

In addition to observing attacker behaviour within these controlled environments, biometric indicators will be utilised to gain deeper insights into cognitive and physiological responses to deception. Metrics such as eye-tracking, pupil dilation, and heart rate variability will be collected to measure stress levels, attentional shifts, and decision hesitancy in response to deceptive stimuli. By integrating both behavioural and psychophysiological indicators, the evaluation process will enable a comprehensive assessment of the model’s descriptive accuracy in capturing real-time adversarial decision-making processes.

The structured nature of these CTF events ensures that data collection is standardised, enabling repeatability and comparability across trials. This approach will not only examine whether the model effectively describes engagement outcomes but also identify potential refinements that enhance its descriptive capabilities. Furthermore, the results will be benchmarked against existing cyber-deception models to illustrate how the proposed model integrates psychological and strategic factors into engagement evaluation.

\subsection{Interpreting the Model}
The proposed model is deliberately abstract, designed to encapsulate the essential cognitive and rational factors that influence an attacker’s decision making and providing a generalised framework for understanding attacker decision-making while remaining adaptable to integration with more complex models. By defining variables such as belief ($B$), scepticism ($S$), experience ($E$), deception fidelity ($D$), and reconnaissance ($R$) within an abstract framework, the model offers a versatile structure that can later be refined with more precise sub-models as needed based on the specific requirements of different cyber-deception frameworks.

This abstraction ensures that individual components could also be replaced or redefined to incorporate data-driven insights from empirical studies or computational models. For example:
\begin{itemize}
\item Belief ($B$) and Scepticism ($S$): These values may initially be represented as simple numerical factors but can be refined using Bayesian belief networks to dynamically update based on newly acquired intelligence or attacker profiling. More complex mathematical representations could incorporate probabilistic graphical models to estimate an attacker’s evolving belief state over time.

\item Deception Fidelity ($D$): The level of deception used in a scenario may be expressed as a discrete value but can be expanded to include machine learning-driven deception quality metrics, where deception efficacy is continuously assessed based on attacker responses. A potential refinement could involve deep learning models that dynamically adjust deception parameters based on observed adversarial behaviour.

\item Reconnaissance ($R$): While currently modelled as a scaling factor, reconnaissance could be linked to real-time network scanning behaviours or social engineering reconnaissance attempts, adapting its weight dynamically based on attacker strategies. More sophisticated formulations could incorporate differential equations to model reconnaissance efforts as a function of time and information gained.

\item Experience ($E$): The impact of experience can be further detailed using reinforcement learning principles, where an attacker’s prior encounters with deceptive systems inform future decision-making tendencies. More advanced approaches could use Markov models to track state-dependent learning effects, capturing long-term behavioural adaptations in adversarial tactics.

\item Engagement ($E_g$): represents the probabilistic outcome describing the extent to which an attacker is likely to interact with a deceptive device when presented with a set of environmental and cognitive conditions. Engagement reflects the combined influence of belief ($B$), scepticism ($S$), deception fidelity ($D$), experience ($E$), and reconnaissance effort ($R$). The resulting value of $E_g$ is bounded by the limits of its input parameters and expresses a relative probability of interaction. Although not restricted to the range of 0 to 1, $E_g$ can be normalised to a bounded probability representation when comparative or cumulative analysis is required. This definition positions engagement as a probabilistic construct derived from the model’s defined relationships, rather than a discrete behavioural count, and allows the value to be expressed consistently across experimental or operational contexts.
\end{itemize}

By maintaining an abstract structure, the proposed model ensures flexibility in its application across various cyber-deception research areas. This adaptability makes it a foundational tool for analysing adversarial engagement without being constrained by rigid parameter definitions. As research advances, the model can be iteratively refined by integrating domain-specific enhancements while preserving its core function as a descriptive decision-evaluation framework.

\subsection{Preliminary Simulation Evaluation}
\label{sec:simulation_evaluation}

The proposed model has not yet been subjected to empirical testing using Capture the Flag (CTF) events. These structured experiments are planned as the next stage of research and will provide the primary empirical validation of the model. At this stage, the evaluation is limited to a preliminary simulation exercise intended to test the internal logic of the model against behavioural patterns reported in prior studies. Monte Carlo simulations were conducted to explore whether the model could reproduce engagement trends consistent with published CTF findings. 

To establish reproducibility and reduce ambiguity in parameter selection, the variable ranges were inferred from published experiments that examined attacker behaviour under deceptive conditions. Table~\ref{tab:parameter_inference} summarises the parameter sources, their empirical basis, and the ranges normalised for use in simulation.

\newcolumntype{Y}{>{\raggedright\arraybackslash}X}

\begin{table}[htbp]
\centering
\small
\caption{Parameter inference and rationale for simulation evaluation}
\label{tab:parameter_inference}
\setlength{\tabcolsep}{3pt} 
\renewcommand{\arraystretch}{1.12}
\begin{tabularx}{\linewidth}{@{}T{2.50cm} T{3.40cm} Y T{2.10cm}@{}}
\toprule
\textbf{Variable} & \textbf{Source of inference} & \textbf{Empirical basis / rationale} & \textbf{Range used} \\
\midrule
Belief (B) &
\textit{Tularosa}~\cite{fergusonwalter2020}, \textit{Moonraker}~\cite{moonraker}, \textit{Nunes}~\cite{nunes} &
Attacker conviction in target authenticity; inferred from engagement frequency or persistence in deceptive scenarios. &
0.6--0.9 \\[0.15em]

Scepticism (S) &
\makecell[l]{\textit{Tularosa}\\informed / \\uninformed~\cite{fergusonwalter2020}} &
Higher scepticism in “informed” conditions reduced engagement; scaled to represent resistance to deception. &
0.1--0.6 \\[0.15em]

Deception fidelity (D) &
\textit{Moonraker}~\cite{moonraker}, \textit{Nunes}~\cite{nunes} &
Apparent realism or complexity of deceptive artefacts; higher values indicate more convincing decoys. &
0.5--0.8 \\[0.15em]

Experience (E) &
\textit{Tularosa}~\cite{fergusonwalter2020}, \textit{Nunes}~\cite{nunes} &
Accumulated familiarity with deception; higher values indicate greater ability to detect false cues. &
0.6--0.9 \\[0.15em]

\makecell[l]{Reconnaissance\\(R)} &
Across studies (scanning and exploration intensity) &
Limits the effects of \textit{E} and \textit{D}; normalised to a continuous scale between 0 and 1, then scaled to study context. &
1.1--1.7 (relative) \\
\bottomrule
\end{tabularx}
\vspace{0.25em}
\footnotesize\emph{Note.} Values are inferred from descriptive statistics and qualitative observations in prior CTF-based research. Ranges initialise Monte Carlo simulations to explore descriptive consistency, not to estimate precise engagement probabilities.
\end{table}

Parameter values derived in Table~\ref{tab:parameter_inference} were then applied in the simulation scenarios summarised in Table~\ref{tab:evaluation_scenarios}. These scenarios correspond to previously published deception experiments—specifically, the Tularosa, Moonraker, and Nunes studies—and were selected to examine the model's ability to reproduce behavioural outcomes under varying deception conditions.

\begin{table}[H]
\centering
\caption{Validation Scenarios with Model Parameters and Outcomes}
\label{tab:evaluation_scenarios}
\renewcommand{\arraystretch}{1.2}
\resizebox{\textwidth}{!}{%
\begin{tabular}{|>{\raggedright\arraybackslash}m{3.5cm}|>{\raggedright\arraybackslash}m{3.2cm}|>{\raggedright\arraybackslash}m{5.1cm}|>{\centering\arraybackslash}m{3.5cm}|}
\hline
\textbf{Study} & \textbf{Parameters Used} & \textbf{Rationale} & \textbf{Graph} \\
\hline
\makecell[l]{Tularosa Study:\\ Uninformed \\Scenario \cite{tularosa}} &
\makecell[l]{$B \sim 0.7$\\ $S \sim 0.4$\\ $D \sim 0.5$\\ $E \sim 0.6$\\ $R \sim 1.1$} &
Participants were unaware of deception. Moderate belief and lower scepticism drove high engagement, as deception fidelity and experience were modest. &
\includegraphics[width=\linewidth]{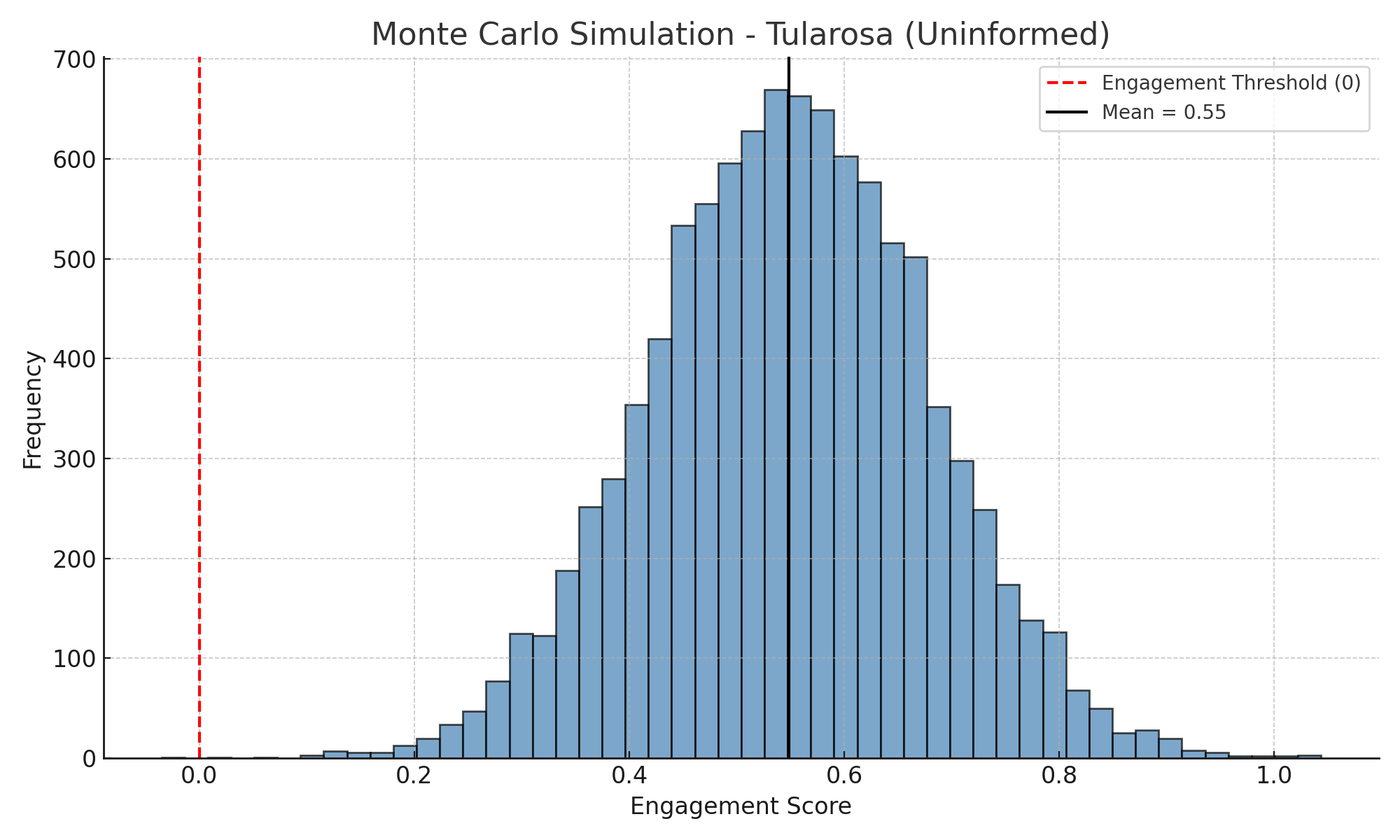} \\
\hline
\makecell[l]{Tularosa Study:\\ Informed Scenario\\\cite{tularosa}} &
\makecell[l]{$B \sim 0.6$\\ $S \sim 0.6$\\ $D \sim 0.7$\\ $E \sim 0.7$\\ $R \sim 1.4$} &
Participants were informed deception was in use. Increased scepticism and experience reduced engagement likelihood. &
\includegraphics[width=\linewidth]{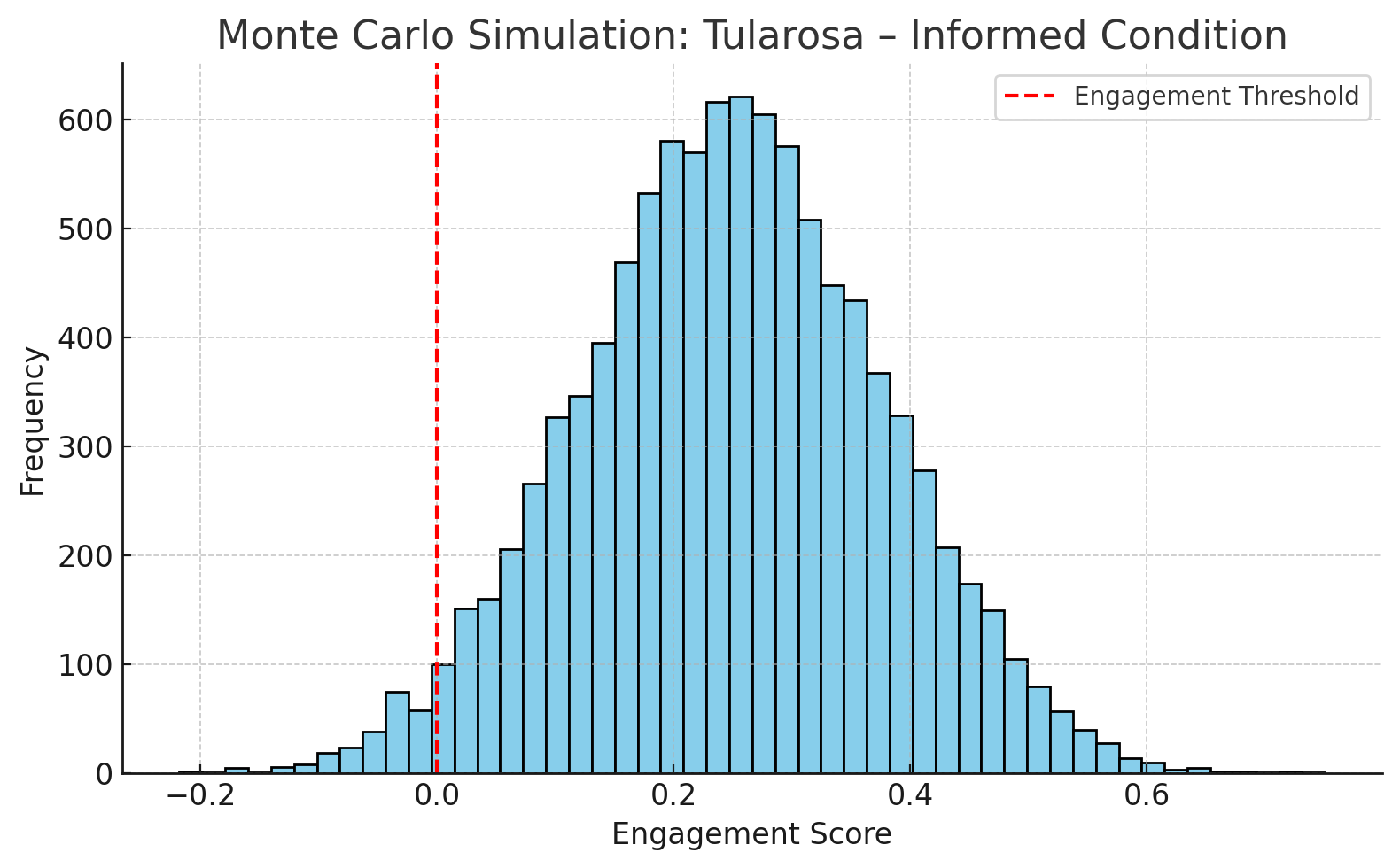} \\
\hline
\makecell[l]{Moonraker Study\\\cite{moonraker}} &
\makecell[l]{$B \sim 0.65$\\ $S \sim 0.3$\\ $D \sim 0.7$\\ $E \sim 0.8$\\ $R \sim 1.5$} &
Deception was host-based and subtle. High belief and fidelity with low scepticism led to increased engagement. &
\includegraphics[width=\linewidth]{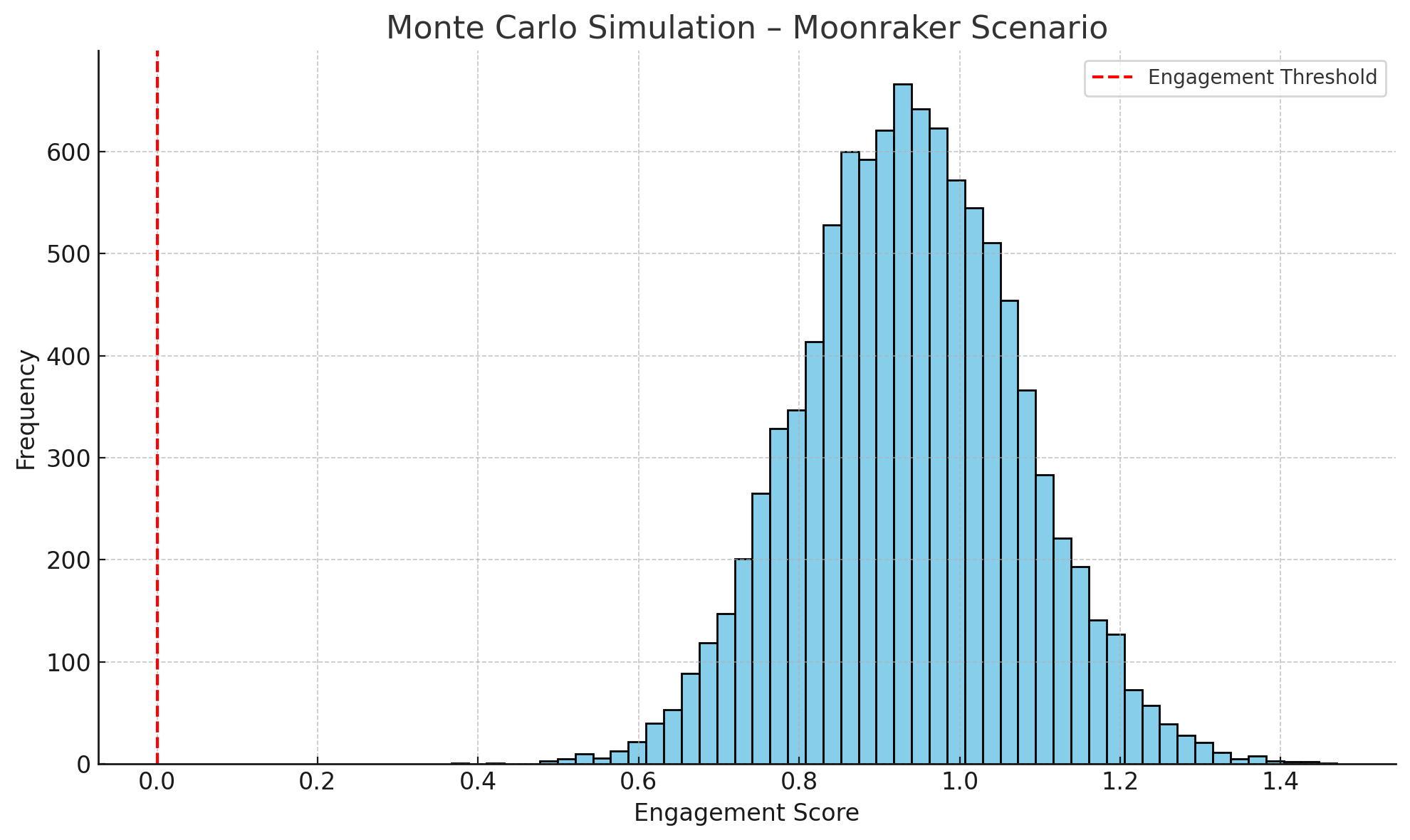} \\
\hline
\makecell[l]{Nunes Study:\\ CTF Attribution \\\& Deception\cite{nunes}} &
\makecell[l]{$B \sim 0.65$\\ $S \sim 0.4$\\ $D \sim 0.8$\\ $E \sim 0.85$\\ $R \sim 1.7$} &
Competitive environment encouraged deception despite awareness. Belief and fidelity outweighed moderate scepticism. &
\includegraphics[width=\linewidth]{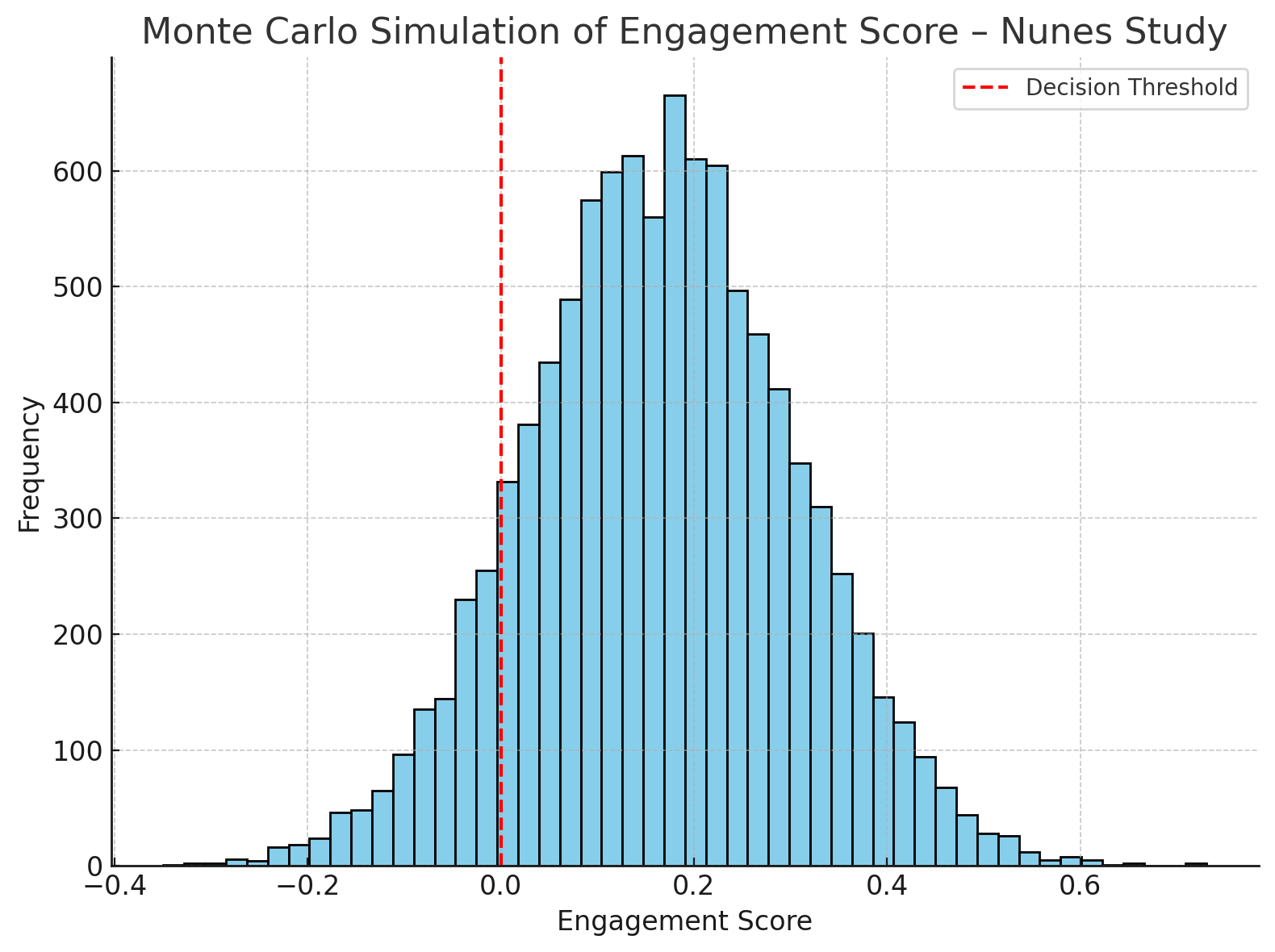} \\
\hline
\makecell[l]{Tularosa Study:\\ Psychological Pilot\\ – No Deception\cite{tularosa}} &
\makecell[l]{$B \sim 0.6$\\ $S \sim 0.5$\\ $D = 0$\\ $E \sim 0.9$\\ $R \sim 1.3$} &
Control group with no deception. High experience and no deceptive cues resulted in complete disengagement. &
\includegraphics[width=\linewidth]{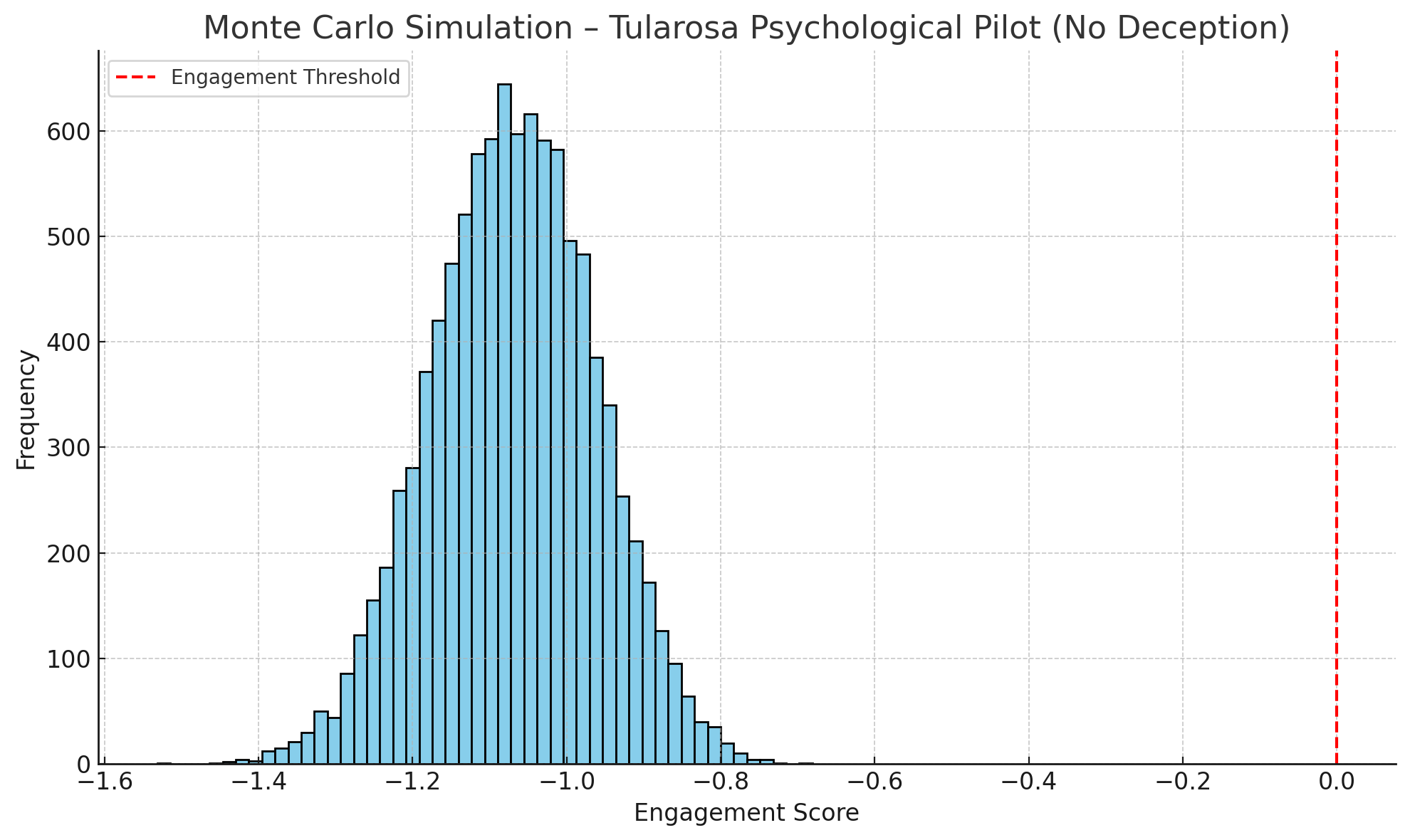} \\
\hline
\end{tabular}%
}
\end{table}

\subsubsection{Tularosa Study – Uninformed Condition}
In the uninformed condition of the Tularosa study, participants were not made aware of the presence of deceptive elements. This scenario was characterised by high belief ($B \sim 0.85$), low scepticism ($S \sim 0.15$), high deception fidelity ($D \sim 0.8$), high experience ($E \sim 0.9$), and elevated reconnaissance effort ($R \sim 1.5$). The model yielded a mean engagement score of $0.55$ with a standard deviation of $0.13$, and $99.99\%$ of simulated attackers chose to engage.

This result aligns with empirical observations reported in the extended Tularosa study by Ferguson-Walter\cite{fergusonwalter2020} which found a statistically significant increase in interactions with deceptive systems under the Present-Uninformed condition. Specifically, the thesis reports that:
\begin{quote}
``A MANOVA found a significant main effect of deception condition on the number of alerts generated $(F(6, 246) = 9.08, p < .001$, partial $\eta^2 = .181$). The decoy scan alerts $(p < .01)$, decoy probe alerts $(p < .001)$, and decoy touch alerts $(p < .001)$ were all significantly higher in the Present-Uninformed condition'' \cite[p.~61]{fergusonwalter2020}.
\end{quote}
These alert types reflect direct engagement with deceptive elements in the environment. Although exact engagement percentages are not provided, the significantly elevated decoy interaction rates suggest that the majority of participants engaged with decoy systems. The model captures this behavioural trend, reflecting how high belief and low scepticism, when deception is not disclosed and appears realistic, result in widespread engagement. The close alignment between the modelled engagement rate and these empirical findings supports the descriptive validity of the proposed model under such conditions.

\subsubsection{Tularosa Study – Informed Condition}
In the informed condition of the extended Tularosa study, participants were explicitly made aware of the presence of deceptive elements. This change in adversarial awareness was reflected in the model parameters as slightly reduced belief ($B \sim 0.75$), increased scepticism ($S \sim 0.35$), and consistent values for deception fidelity ($D \sim 0.8$), experience ($E \sim 0.9$), and reconnaissance ($R \sim 1.5$). The resulting Monte Carlo simulation yielded a mean engagement score of $0.25$, with a standard deviation of $0.13$, and an engagement rate of $97.08\%$.

These outcomes align with empirical findings reported by Ferguson-Walter (2020), who observed:
\begin{quote}
``Despite being warned about the deception, participants in the Present-Informed group still engaged with decoys at high rates'' \cite[p.~84]{fergusonwalter2020}.
\end{quote}
The model effectively reflects this behavioural shift. Compared to the uninformed condition, where belief remained high and scepticism low, the informed group demonstrated increased caution. This is quantitatively captured through a reduced engagement score and is consistent with the observed behavioural modulation in the study. While participants remained active in the environment, the altered engagement profile demonstrates that the proposed model can sensitively reflect the influence of deception awareness on adversarial decision-making.

\subsubsection{Moonraker Scenario – Uninformed Condition}
In the Moonraker study, attackers were unaware of deceptive elements, resulting in higher engagement with decoy assets and reduced task performance. To model this condition, parameter values were inferred as follows: very high belief ($B \sim 0.9$), low scepticism ($S \sim 0.1$), high deception fidelity ($D \sim 0.75$), moderate experience ($E \sim 0.6$), and high reconnaissance effort ($R \sim 1.6$). These values reflect a condition in which attackers encountered believable decoys without prior knowledge that deception was in play, and where insufficient experience with deceptive devices allowed high-fidelity deception to succeed.

A Monte Carlo simulation was conducted with 10,000 trials using these parameters. The resulting engagement score had a mean of $0.94$, a standard deviation of $0.14$, and a simulated engagement rate of $100.00\%$.

These results are consistent with the findings of Shade et al.\cite{moonraker}, who reported:
\begin{quote}
``Participants took longer to complete the scenario when deception was present, suggesting they were engaging with the decoy content'' \cite[p.~3]{moonraker}.
\end{quote}

Although the study does not directly report a percentage of engagement with deceptive elements, the extended task times and decreased effectiveness among participants strongly imply substantial interaction with decoys. The proposed model reflects this behavioural outcome, showing that a combination of high belief and low scepticism, when not sufficiently counterbalanced by experience, results in widespread engagement with deception. The simulated results thus align well with the observed data.

These outcomes are highly consistent with experimental results, in which the absence of awareness led to strong engagement with deceptive elements. The proposed model strongly aligns with the study's findings by attributing high engagement to the overpowering influence of belief and low scepticism, especially when experience is insufficient to offset deception.

\subsubsection{Nunes Study – CTF Attribution and Deception}
The study by Nunes et al.\cite{nunes} investigated adversarial behaviour and deceptive strategies during the DEFCON 21 Capture-the-Flag (CTF) competition. Operating in a hostile environment where deception was both expected and strategically beneficial, participants demonstrated aggressive tactics despite known risks of counter-deception. To model this environment, parameter values were inferred based on qualitative descriptions of attacker conduct: belief ($B \sim 0.65$), scepticism ($S \sim 0.4$), deception fidelity ($D \sim 0.8$), experience ($E \sim 0.85$), and reconnaissance effort ($R \sim 1.7$). The Monte Carlo simulation produced a mean engagement score of $0.16$, a standard deviation of $0.12$, and an engagement rate of $87.62\%$.

This result closely aligns with the observed behaviour described in the study. Nunes et al.\cite{nunes} report that:
\begin{quote}
``Players frequently launched attacks even when attribution could not be assured and deception was highly likely. Despite understanding the risks, participants proceeded with offensive actions to secure points or disrupt opponents.''
\end{quote}
This assertion shows that participants remained actively engaged, even under conditions of known risk and strategic uncertainty. The model reflects this behavioural trend by demonstrating how belief in a target’s value, when combined with high attacker experience, can override moderate scepticism, particularly in environments where deception is a normative component of the engagement. The alignment between the model’s output and the study’s documented outcomes reinforces the descriptive validity of the proposed model in competitive, deception-aware CTF scenarios.

\subsubsection{Tularosa Study: Psychological Pilot – No Deception}
This control condition from the Tularosa study involved no deceptive elements. Parameters were inferred to reflect cautious but neutral engagement tendencies: moderate belief ($B \sim 0.6$), moderate-to-high scepticism ($S \sim 0.5$), no deception ($D = 0$), high experience ($E \sim 0.9$), and moderate reconnaissance ($R \sim 1.3$). The resulting Monte Carlo simulation yielded a mean engagement score of $-1.07$, a standard deviation of $0.11$, and an engagement rate of $0.00\%$.

These simulation results align with empirical findings reported in Ferguson-Walter’s extended Tularosa thesis, which observed:

\begin{quote}
``Participants in the control condition had significantly fewer decoy alerts and lower deception-related activity compared to those in deception-present conditions, indicating little or no engagement with deceptive elements''\cite{tularosa}.
\end{quote}

This output validates the model’s ability to reflect disengagement in the absence of deceptive signals, particularly when scepticism and experience levels are high and belief in target value is moderate.

\subsubsection{Summary of Simulation Results}
 \begin{table}[H]
    \centering
    \caption{Summary of Simulation Results}
    \begin{tabular}{lccc}
    \toprule
    \textbf{Scenario} & \textbf{Mean Score} & \textbf{Std Dev} & \textbf{\% Engaged} \\
    \midrule
    Tularosa – Uninformed & 0.55 & 0.13 & 99.99\% \\
    Tularosa – Informed & 0.25 & 0.13 & 97.08\% \\
    Moonraker – With Deception & 0.94 & 0.14 & 100.00\% \\
    Nunes – CTF Attribution & 0.16 & 0.12 & 87.62\% \\
    Tularosa – Psychological Pilot & -1.07 & 0.11 & 0.00\% \\
    \bottomrule
    \end{tabular}
\end{table}

Across all scenarios, the model behaved consistently with the behavioural expectations informed by prior empirical work. The uninformed Tularosa condition produced the highest engagement rate (99.99\%), reflecting a combination of high belief, low scepticism, and credible deceptive cues. When deception was disclosed in the informed Tularosa condition, engagement decreased to 97.08\%, demonstrating a measurable but limited effect of increased scepticism. 

The Moonraker scenario yielded a mean engagement score of 0.94 with 100.00\% engagement, aligning with the high levels of interaction reported when deceptive elements were both believable and undisclosed. In contrast, the Nunes scenario, characterised by higher scepticism and competitive adversarial pressure, resulted in a lower engagement rate of 87.62\%, indicating that elevated uncertainty and experience can moderate engagement without eliminating it. 

The psychological pilot condition, which contained no deception, produced a negative mean engagement score and 0.00\% engagement. This outcome reflects circumstances where scepticism outweighs belief and no deceptive cues are present to influence attacker perception. 

Overall, these results demonstrate that the model responds predictably to variations in belief, scepticism, deception fidelity, experience, and reconnaissance. The engagement outcomes scale in a manner that is coherent with the behavioural expectations inferred from the empirical studies referenced in Section~4.2, supporting the descriptive utility of the model at this early stage.

\subsubsection{Interpreting the Model Output}
The model produces an engagement result ($E_g$) that represents the relative likelihood that an attacker will interact with a deceptive device. During simulation, $E_g$ is calculated as a bounded value determined by the limits of its input parameters. For clarity of presentation, this bounded result is expressed as a normalised probability, $P(E_g)$, which is reported in Table 4 as the probability of engagement shown as a percentage. The magnitude and sign of $E_g$ describe the direction and strength of the influence exerted by belief ($B$), scepticism ($S$), deception fidelity ($D$), experience ($E$), and reconnaissance effort ($R$), while $P(E_g)$ provides a standardised measure that facilitates direct comparison between conditions.

This dual representation distinguishes the model’s internal computation from its interpretive output: $E_g$ defines the bounded relationship among variables, and $P(E_g)$ communicates that relationship through a conventional probabilistic form. Presenting results in this manner maintains consistency with the model’s theoretical basis while enabling clear interpretation of outcomes across different parameter configurations. It also allows $P(E_g)$ to be directly compared with empirical observations when available, linking simulated outcomes with measurable attacker behaviour without constraining the model to a specific probability scale.

\subsubsection{Sensitivity and Robustness Discussion}
Preliminary observations from the simulation indicate that variations in the input parameters influence the engagement result ($E_g$) in predictable ways. Because $E_g$ is a bounded probabilistic outcome, changes in belief ($B$) or deception fidelity ($D$) generally increase engagement, while equivalent increases in scepticism ($S$) or experience ($E$) tend to reduce it. Reconnaissance effort ($R$) moderates these effects by influencing the relative strength of each component, ensuring that no single variable dominates the outcome outside its defined range. These relationships demonstrate that the model behaves consistently with the conceptual expectations of attacker decision-making and that changes in $E_g$ reflect coherent behavioural trends rather than arbitrary parameter interactions.

Although the present stage of development does not include a dedicated sensitivity analysis, the simulation behaviour suggests that the model is internally stable and mathematically coherent. The observed relationships between parameters align with the underlying theoretical rationale, indicating that the model responds proportionally to changes in its inputs. Future work will extend this evaluation through structured sensitivity testing and empirical calibration, allowing the relative influence of each parameter to be validated against observed attacker behaviour. This will provide a stronger empirical basis for assessing the model’s robustness and generalisability across different deceptive environments.

\section{Conclusion}
This paper introduced a descriptive model for understanding attacker decision-making in deceptive cyber environments. The model formalises engagement as a bounded probabilistic outcome influenced by belief, scepticism, deception fidelity, experience, and reconnaissance, providing an interpretable structure for examining how cognitive factors shape the decision to interact with a deceptive device. By representing these factors explicitly, the model contributes a conceptual foundation for exploring attacker perception prior to active engagement.

A preliminary simulation evaluation was undertaken to examine how variations in these parameters influence the engagement outcome. The simulation results aligned with the expected behavioural tendencies informed by the empirical studies referenced in Section~4.2, indicating that the model behaves consistently with theoretical expectations at this early stage of development. These observations demonstrate that the model produces coherent and predictable responses to parameter variation, allowing descriptive exploration of attacker behaviour across differing deceptive contexts. However, these findings remain indicative rather than confirmatory and do not constitute empirical validation.

The model currently retains several limitations typical of early-stage conceptual frameworks. Its abstract representation means that parameter ranges are coarse and uncalibrated, and the relationships between components have yet to be validated against behavioural data. Although the preliminary simulation indicates internal coherence, comprehensive sensitivity analysis and empirical calibration are required before the model can be applied predictively or operationally. These limitations reflect the developmental stage of the framework rather than deficiencies in its conceptual basis.

Future research will focus on empirical validation through structured capture-the-flag exercises designed to elicit attacker perceptions and behavioural cues in controlled deceptive environments. Such experiments will support the refinement of parameter ranges, the calibration of engagement outputs, and the integration of the model with complementary computational approaches, including Bayesian inference, Markov decision processes, and reinforcement learning. Advancing this line of research has the potential to enhance the design and evaluation of cyber-deception strategies by providing a tractable and conceptually grounded framework for understanding attacker engagement within complex defensive settings.

\newpage
\appendix
\section{References}
\label{sec:References:appendix}
 \bibliographystyle{elsarticle-num} 
 \bibliography{cas-refs.bib}

@ARTICLE{Benke2021,
  author  = {Benke, L. and Papasimeon, M. and Miller, T.},
  title   = {Modelling strategic deceptive planning in adversarial multi-agent systems.},
  journal = {Communications in Computer and Information Science},
  year    = {2021},
  url     = {https://doi.org/10.48550/arxiv.2109.03092}
}

@ARTICLE{Black2020,
  author  = {Black, R. and Reid, I.},
  title   = {Toward a holistic model of deception: subject matter expert validation.},
  journal = {Proceedings of the Annual Hawaii International Conference on System Sciences.},
  year    = {2020},
  url     = {https://doi.org/10.21236/ada426011}
}

@ARTICLE{Edwards2004,
  author  = {Edwards, R.},
  title   = {Allies in the shadows: why we need operational deception.},
  journal = {Naval War College},
  year    = {2004},
  url     = {https://doi.org/10.24251/hicss.2020.230}
}

@ARTICLE{Grohe2007,
  author  = {Grohe, E.},
  title   = {Military deception: transparency in the information age.},
  journal = {Naval War College},
  year    = {2007},
  url     = {https://doi.org/10.21236/ada476638}
}

@ARTICLE{Hood2023,
  author  = {Hood, D.},
  title   = {Deterrence by deception: historic lessons and contemporary methods.},
  journal = {Contemporary Issues in Air and Space Power.},
  year    = {2023},
  url     = {https://doi.org/10.58930/001c.84546}
}

@ARTICLE{Knobloch2020,
  author  = {Knobloch, L. and Knobloch‐Fedders, L. and Yorgason, J. and Basinger, E. and Abendschein, B. and McAninch, K.},
  title   = {Suspicion about a partner’s deception and trust as roots of relational uncertainty during the post-deployment transition.},
  journal = {Journal of Social and Personal Relationships},
  year    = {2020},
  pages   = {912-934},
  url     = {https://doi.org/10.1177/0265407520970645}
}

@ARTICLE{moonraker,
  author  = {Shade, T. and Rogers, A. and Ferguson‐Walter, K. and Elsen, S. and Fayette, D. and Heckman, K.},
  title   = {The moonraker study: an experimental evaluation of host-based deception.},
  journal = {Proceedings of the 53rd Hawaii International Conference on System Sciences (HICSS)},
  year    = {2020},
  pages   = {1-10},
  url     = {https://doi.org/10.24251/hicss.2020.231}
}

@ARTICLE{tularosa,
  author  = {Ferguson‐Walter, K. and Shade, T. and Rogers, A. and Trumbo, M. C. S. and and Nauer, K. S. and Divis, K. M. and Jones, A. and Combs, A. and Abbott, R. G.},
  title   = {The Tularosa Study: An Experimental Design and Implementation to Quantify the Effectiveness of Cyber Deception.},
  journal = {Proceedings of the 52nd Hawaii International Conference on System Sciences.},
  year    = {2019},
  pages   = {874-883},
  url     = {https://www.osti.gov/servlets/purl/1524844}
}

@ARTICLE{nunes,
  author  = {Nunes, E. and Kulkarni, N. and Shakarian, P. and Ruef, A. and Little, J.},
  title   = {Cyber-deception and attribution in capture-the-flag exercises.},
  journal = {Proceedings of the 2015 IEEE/ACM International Conference on Advances in Social Networks Analysis and Mining 2015},
  year    = {2015},
  pages   = {962-965},
  url     = {https://dl.acm.org/doi/abs/10.1145/2808797.2809362?casa_token=gEkhyOB3BaMAAAAA:K1ICVrQoUK84VI0fIfIZnD6zCor64Xkd2tN-aJPw7SnvkruODDvOQ0ORW1TMugszcY50JKLRuUaU1TE}
}

@phdthesis{fergusonwalter2020,
  author       = {Ferguson-Walter, Kimberly J.},
  title        = {An Empirical Assessment of the Effectiveness of Deception for Cyber Defense},
  school       = {Naval Postgraduate School},
  year         = {2020},
  type         = {PhD thesis},
  url          = {https://apps.dtic.mil/sti/pdfs/AD1115369.pdf}
}
\end{document}